\documentclass[a4paper]{jpconf}
\usepackage{graphicx}
\usepackage{caption}
\usepackage{subcaption}
\usepackage{citesort}
\usepackage{xcolor}

\usepackage{lineno}

\begin{document}
\title{Hyperparameter optimization of data-driven AI models on HPC systems}

\author{Eric Wulff$^1$, Maria Girone$^1$ and Joosep Pata$^2$}

\address{$^1$CERN, Esplanade des Particules 1, 1211 Geneva 23, Switzerland}
\address{$^2$NICPB, Rävala pst 10, 10143 Tallinn, Estonia}

\ead{eric.wulff@cern.ch}

\begin{abstract}
In the European Center of Excellence in Exascale Computing "Research on AI- and Simulation-Based Engineering at Exascale" (CoE RAISE), researchers develop novel, scalable AI technologies towards Exascale. This work exercises High Performance Computing resources to perform large-scale hyperparameter optimization using distributed training on multiple compute nodes. This is part of RAISE’s work on data-driven use cases which leverages AI- and HPC cross-methods developed within the project.
In response to the demand for parallelizable and resource efficient hyperparameter optimization methods, advanced hyperparameter search algorithms are benchmarked and compared. The evaluated algorithms, including Random Search, Hyperband and ASHA, are tested and compared in terms of both accuracy and accuracy per compute resources spent.
As an example use case, a graph neural network model known as MLPF, developed for Machine Learned Particle-Flow reconstruction, acts as the base model for optimization. Results show that hyperparameter optimization significantly increased the performance of MLPF and that this would not have been possible without access to large-scale High Performance Computing resources. It is also shown that, in the case of MLPF, the ASHA algorithm in combination with Bayesian optimization gives the largest performance increase per compute resources spent out of the investigated algorithms.

\end{abstract}

\section{Introduction}
One of the primary goals in the European Center of Excellence in Exascale Computing “Research on AI- and Simulation-Based Engineering at Exascale” (CoE RAISE) \cite{raise_website} is the development and expansion of Artificial Intelligence (AI) and High-Performance Computing (HPC) methods along representative use cases from research and industry. While Work Package 3 (WP3) “Compute-Driven Use-Cases at Exascale” covers use cases that are compute-driven, WP4 “Data-Driven Use-Cases at Exascale” has a strong focus on data-driven technologies, i.e., analyzing data-rich descriptions of physical phenomena. Example use cases vary widely and range from fundamental physics and remote sensing to 3D printing and acoustics.

The work of WP4 is highly integrated with WP2 “AI- and HPC-Cross Methods at Exascale”. Experts in WP2 provide support on HPC and AI methods to use cases in WP4. This support manifests itself in porting code to new HPC architectures and machines, in performance analyses and engineering of codes, and in the development of AI solutions for the individual use cases.

In the work presented here, HPC resources are leveraged to perform large-scale hyperparameter optimization (HPO) using distributed training on multiple nodes as part of WP4. As an example use case from the field of High Energy Physics (HEP), the AI-based particle-flow reconstruction algorithm called Machine-Learned Particle-Flow (MLPF) \cite{Pata:2021oez} acts as the base model for which HPO is performed. MLPF is developed in the Compact-Muon-Solenoid (CMS) Collaboration \cite{Collaboration_2008} at CERN and combines information from tracks and calorimeter clusters to reconstruct particle candidates.

Further developments of AI in RAISE have the potential to greatly impact the field of High Energy Physics by efficiently processing the large amounts of data that will be produced by particle detectors in the coming decades. Moreover, HPO is model agnostic and could be widely applied in other sciences using AI, e.g., in the fields of seismic imaging, remote sensing, defect-free additive manufacturing and sound engineering that are part of Work Package 4.


HPO, sometimes referred to as hyperparameter tuning or hypertuning, is the process of tuning the hyperparameters of the model to optimize its performance. Hyperparameters are the parameters that are not learned during the model training but must be defined by the user. Some examples are model-architecture-related parameters such as the number of layers in the model and the number of nodes in each layer, or optimization-related parameters such as the batch size and the learning rate.

Hypertuning deep learning-based AI models is often compute resource intensive, partly due to the high cost of training a single hyperparameter configuration to completion and partly because of the infinite set of possible hyperparameter combinations to evaluate. There is therefore a need for large-scale, parallelizable and resource efficient hyperparameter search algorithms.

This work makes use of a distributed computing tool called Ray \cite{DBLP:journals/corr/abs-1712-05889}, and more specifically the part of Ray called Tune \cite{DBLP:journals/corr/abs-1807-05118}. Tune is an open-source tool for multi-node distributed hypertuning which integrates well with modern machine learning frameworks like e.g., TensorFlow \cite{tensorflow2015-whitepaper} and PyTorch \cite{NEURIPS2019_9015}. It also supports integration with many other hypertuning tools such as Scikit-Optimize \cite{scikit-optimize}, HyperOpt \cite{pmlr-v28-bergstra13}, Optuna \cite{DBLP:journals/corr/abs-1907-10902}, SigOpt \cite{sigopt-web-page}, and more.

In the following, section \ref{sec:cern} describes the example use case for which HPO is performed, section \ref{sec:raise} describes how HPO can be used in a wide variety of applications and highlight synergies within CoE RAISE, and finally, section \ref{sec:conclusion} presents the conclusions.

\section{ Example use case: Event reconstruction and classification at the CERN HL-LHC}
\label{sec:cern}


With the upcoming upgrade of the Large Hadron Collider (LHC) to the High Luminosity LHC (HL-LHC), the HEP community
will face a significant  
increase in data production.
This  
motivates efforts to optimize the speed and efficiency with which data is collected, processed, and analyzed, and is one of the major challenges that must be solved by the time the HL-LHC starts operation at the end of 2027.

One of the many different approaches that are being investigated to tackle this challenge is to replace traditional HEP algorithms with faster, parallelizable AI-driven approaches. These approaches promise to deliver similar or even better physics performance and can relatively easily be accelerated by hardware such as Graphics Processing Units (GPUs) or Field Programmable Gate Arrays (FPGAs).

One such traditional algorithm that could potentially be replaced by an AI-based version is the so-called Particle-Flow (PF) reconstruction algorithm \cite{Sirunyan_2017}. It processes signals from different sub-detectors and combines them to construct higher-level physics objects. These objects are used for downstream workflows and are important for physics analyses involving hadronic jets and missing transverse energy. An effort to construct a machine-learned PF algorithm is the so-called MLPF algorithm, which is based on a deep neural network implemented using a Graph Neural Network (GNN) formalism. A detailed description of its first iteration can be found in \cite{Pata:2021oez} while a more recent version is described in \cite{mlpf-proceedings}. The code to build, train, and evaluate the model is publicly available online~\cite{joosep_pata_2021_5520559}.

The best performing MLPF hyperparameters were found after two stages of hypertuning. The first stage was performed on the Jülich Wizard for European Leadership Science (JUWELS) Booster 
\cite{JUWELS} at the Jülich Supercomputer Centre in  Jülich, Germany, and required 
19,574 core-hours to complete.
Each compute node on the JUWELS Booster has two AMD EPYC Rome 7402 CPUs with 48 cores clocked at 2.8 GHz and four NVIDIA A100-SXM4-40GB GPUs. The so-called Bayesian Optimization Hyperband (BOHB) \cite{DBLP:journals/corr/abs-1807-01774} algorithm was used to tune parameters of the optimizer such as the \texttt{lr} and the learning rate schedule as well as the \texttt{dropout} and other model-specific internal hyperparameters. The BOHB search space is summarized in table \ref{tab:bohb_searchspace}.

The second hypertuning stage was performed on CoreSite at the Flatiron Institute in New York, NY, USA, using twelve compute nodes, each equipped with a 64-core Intel Icelake Platinum 8358 CPU clocked at 2.6 GHz and four NVIDIA A100-SXM4-40GB GPUs. The best hyperparameter values found from the first search were fixed and stage two instead tuned various architecture parameters such as the number of graph layers and the number of graphs in each layer, as well as the number of nodes and layers used for decoding, and a few other model-specific parameters. The search space of the second stage is summarized in table \ref{tab:asha_searchspace}. In addition, a different hypertuning algorithm called Asynchronous Successive Halving Algorithm (ASHA) \cite{DBLP:journals/corr/abs-1810-05934} was used in combination with Bayesian optimization. The ASHA algorithm allows for an efficient use of compute resources when performing distributed multi-node hypertuning by early stopping trials that underperform relative to others. The second stage of hypertuning consumed approximately 56,730 core-hours. This work would not have been possible without access to HPC resources, as can be illustrated by a back-of-the-envelope calculation to compute that the two hypertuning stages would have taken roughly 6 months to complete using a single GPU, compared to about 83 hours using supercomputers.

\begin{table}[htb]
\begin{minipage}[b]{0.49\linewidth}
	\centering
	\caption{Search space used in the hypertuning run using the BOHB algorithm.}
	\vspace{-2mm}
	\label{tab:bohb_searchspace}
	\footnotesize
		\begin{tabular}{ll}
			\br
			Hyperparameter                              & Search space \\
			\mr
			\texttt{lr}                     &  $\log{lr} \sim \textsc{U}(10^{-4}, 3\cdot10^{-2}))$ \\
			\texttt{dropout}                & (0, 0.5) \\
			\texttt{clip\_value\_low}         &  (0, 0.2) \\
			\texttt{dist\_mult}              &  (0.01, 0.2) \\
			\br
		\end{tabular}
		\vspace{-5mm}
\end{minipage}
\quad
\begin{minipage}[b]{0.49\linewidth}
	\centering
	\caption{Search space used for ASHA in combination with Bayesian optimization.}
	\label{tab:asha_searchspace}
	\vspace{-2mm}
	\footnotesize
		\begin{tabular}{ll}
			\br
			Hyperparameter                      & Search space \\
			\mr
			\texttt{bin\_size}                  &  \{16, 32, 40, 64, 80\} \\
			\texttt{distance\_dim}              &  \{32, 64, 128, 256\} \\
			\texttt{ffn\_dist\_hidden\_dim}     & \{32, 64, 128, 256\} \\
			\texttt{ffn\_dist\_num\_layers}     &  \{1, 2, 3, 4\} \\
			\texttt{num\_graph\_layers\_common} &  \{1, 2, 3, 4\} \\
			\texttt{num\_graph\_layers\_energy} &  \{1, 2, 3, 4\} \\
			\texttt{num\_node\_messages }       &  \{1, 2, 3, 4\} \\
			\texttt{output\_dim}                &  \{32, 64, 128, 256\} \\
			\br
		\end{tabular}
\end{minipage}
\end{table}

\begin{figure}[h]
	\includegraphics[width=0.7\linewidth]{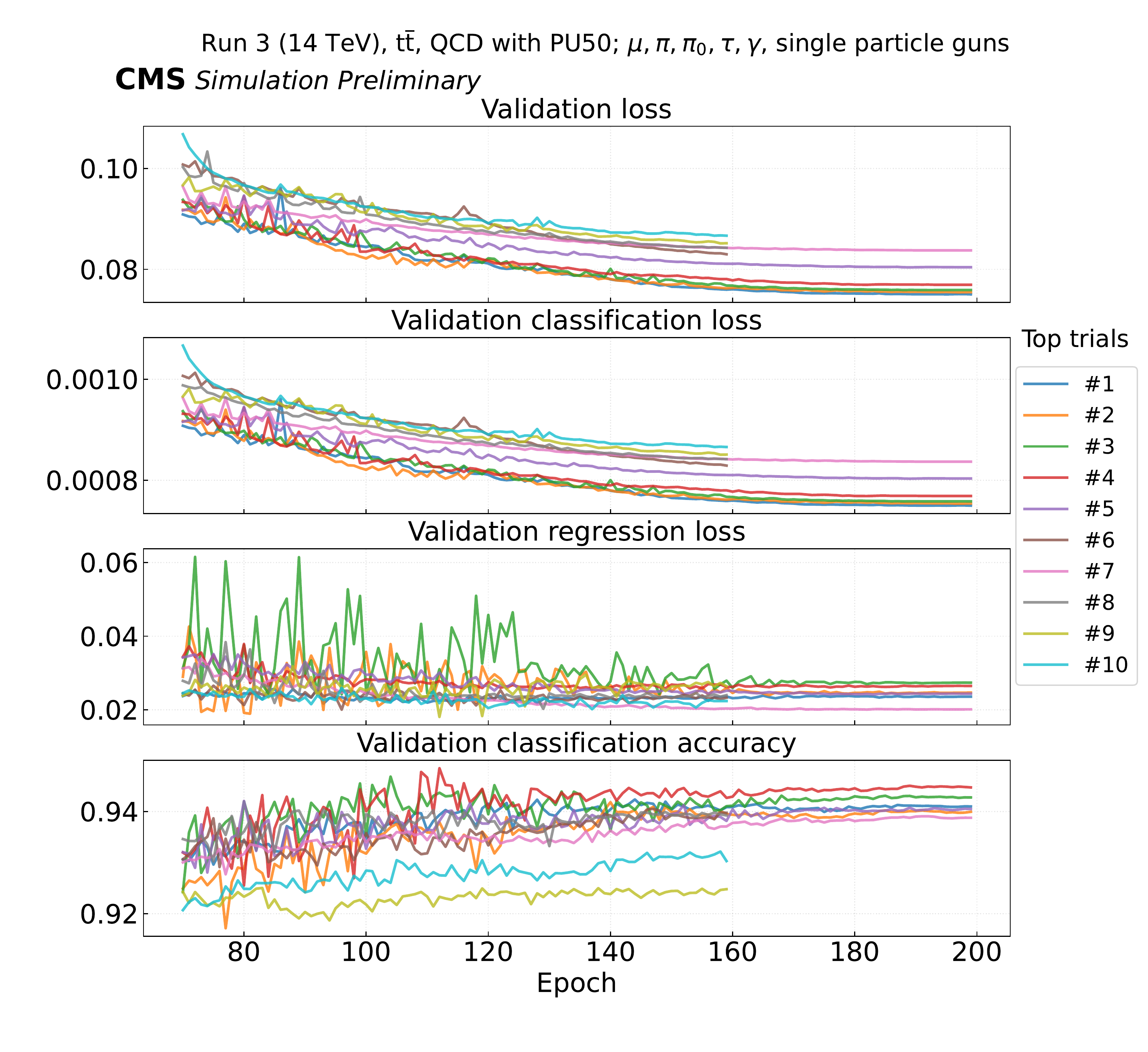}\hspace{2mm}
	\begin{minipage}[b]{0.28\linewidth} \caption{Loss and accuracy curves for top performing trials according to validation loss after hypertuning using the ASHA algorithm in combination with Bayesian optimization drawing 200 samples from the search space. From top to bottom: validation loss, validation classification loss, validation regression loss and validation classification weighted accuracy. The trials were trained for up to 200 epochs but the plots zoom in on epochs 70 and onward for better visibility.}
		\label{fig:asha-plots}
	\end{minipage}
\end{figure}

In both stages described above, the search algorithms were allowed to draw 200 samples from the hyperparameter search space. The best hyperparameters 
found
according to validation loss after both stages of hypertuning are reported in table
\ref{tab:best}
and various metrics as a function of the training epoch are shown in figure \ref{fig:asha-plots}.

\begin{table}[htb]
	\centering
	\caption{Best hyperparameters found.}
	\vspace{-2mm}
	\label{tab:best}
	\footnotesize
	\lineup
		\begin{tabular}{ll}
			\br
			Hyperparameter                              & Value \\
			\mr
			\texttt{lr}                     & \0\00.001129 \\
			\texttt{dropout}                & \0\00.016312 \\
			\texttt{clip\_value\_low}         &  \0\00.001998 \\
			\texttt{dist\_mult}              &  \0\00.120898 \\
			\texttt{bin\_size}                  &  \064 \\
			\texttt{distance\_dim}              &  \064 \\
			\texttt{ffn\_dist\_hidden\_dim}     & 128 \\
			\texttt{ffn\_dist\_num\_layers}     &  \0\03 \\
			\texttt{num\_graph\_layers\_common} &  \0\03 \\
			\texttt{num\_graph\_layers\_energy} &  \0\02 \\
			\texttt{num\_node\_messages }       &  \0\03 \\
			\texttt{output\_dim}                &  \064 \\
			\br
		\end{tabular}
\end{table}

\begin{figure}[htb]
	\centering
	\begin{subfigure}[b]{0.495\textwidth}
		\centering
		\includegraphics[width=1.0\linewidth]{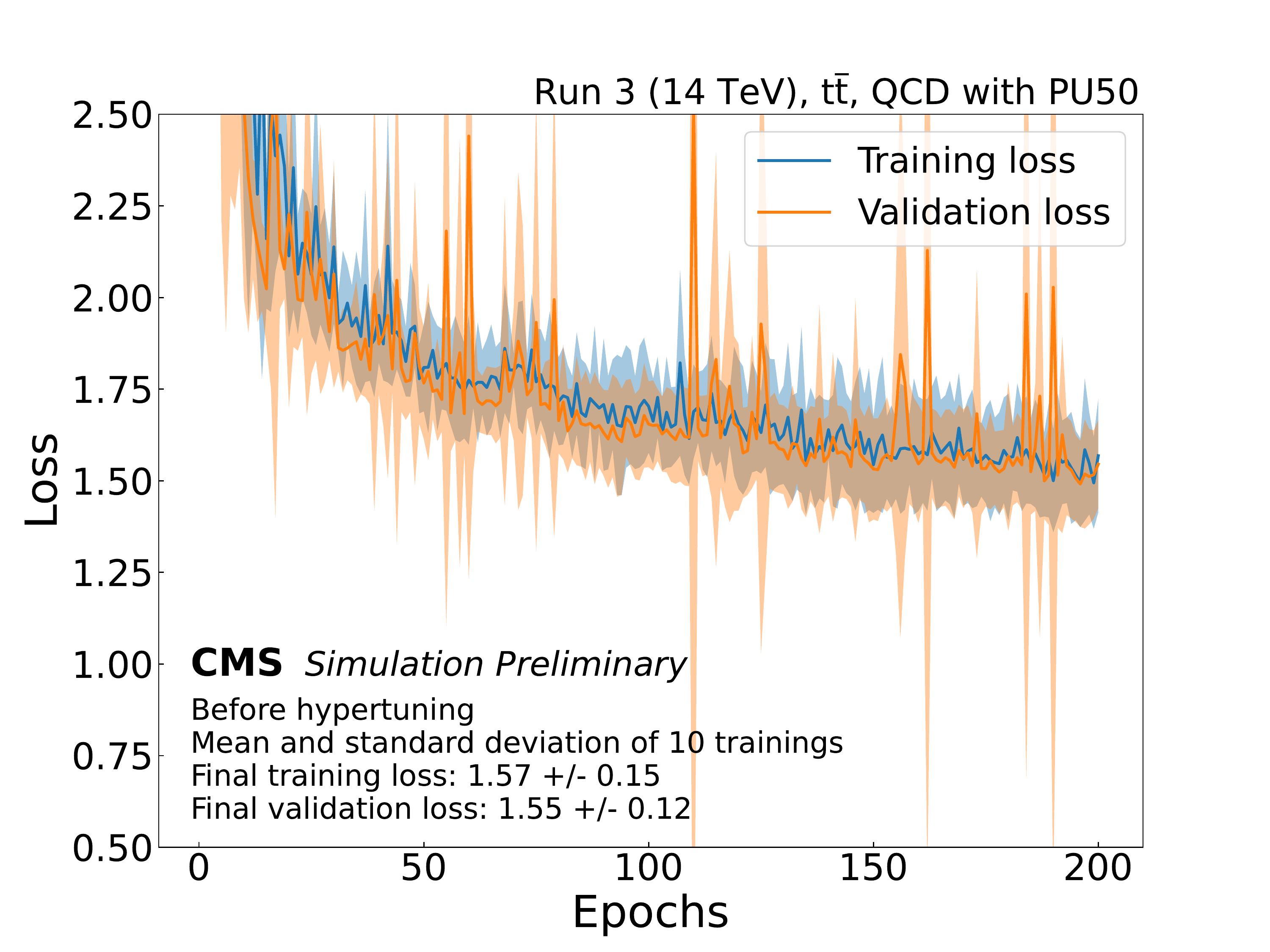}
		\\ \vspace{-3mm}
		\caption{Loss curves before hypertuning.}
		\label{fig:loss-before}
	\end{subfigure}
	\begin{subfigure}[b]{0.495\textwidth}
		\centering
		\includegraphics[width=1.0\linewidth]{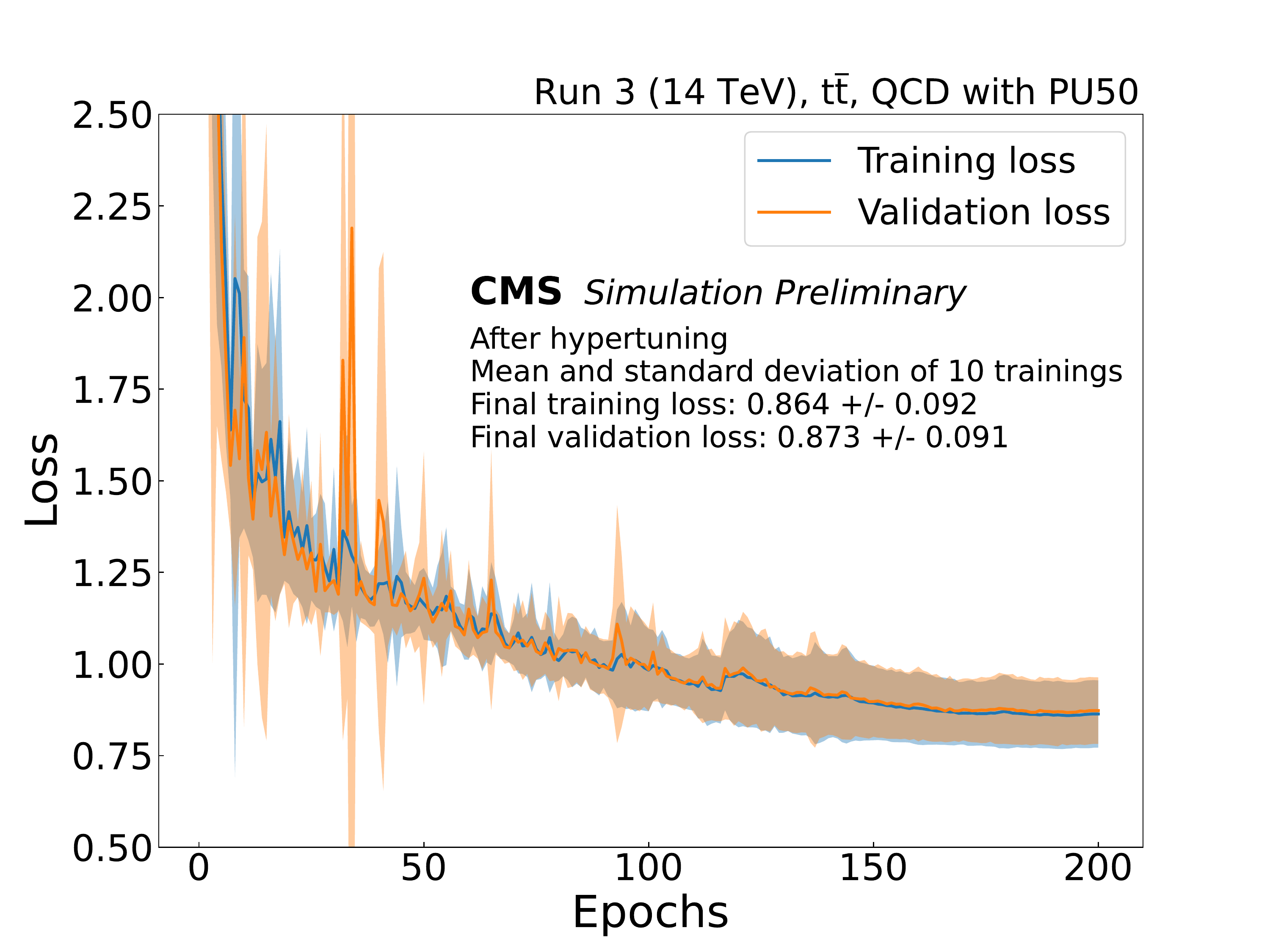}
		\\ \vspace{-3mm}
		\caption{Loss curves after hypertuning}
		\label{fig:loss-after}
	\end{subfigure}
\\ \vspace{-1mm}
	\begin{subfigure}[b]{0.495\textwidth}
		\centering
		\includegraphics[width=1.0\linewidth]{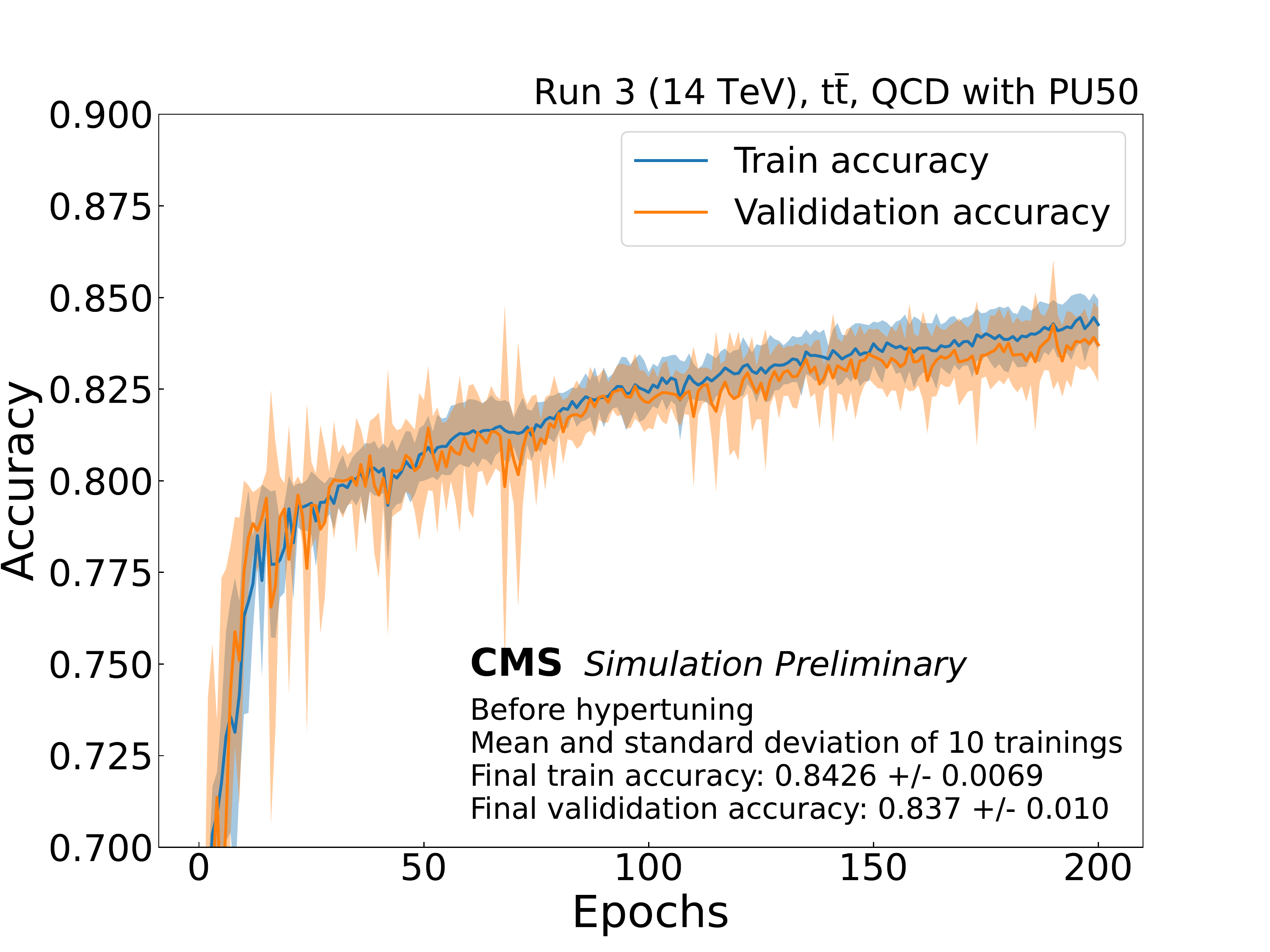}
		\\ \vspace{-3mm}
		\caption{Accuracy curves before hypertuning.}
		\label{fig:acc-before}
	\end{subfigure}
	\begin{subfigure}[b]{0.495\textwidth}
		\centering
		\includegraphics[width=1.0\linewidth]{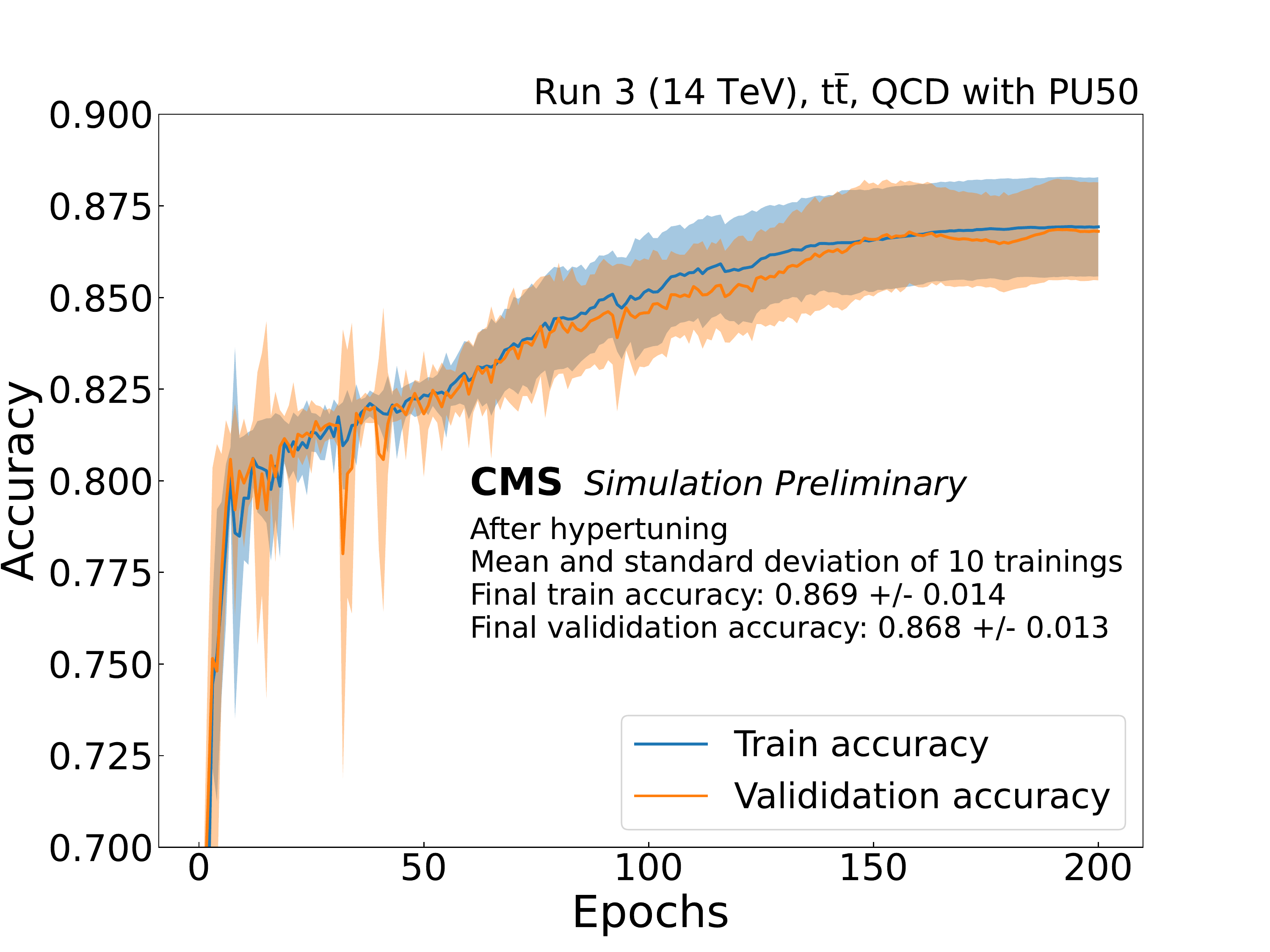}
		\\ \vspace{-3mm}
		\caption{Accuracy curves after hypertuning.}
		\label{fig:acc-after}
	\end{subfigure}
	\\ \vspace{-2mm}
	\caption{Mean and standard deviation of loss and classification accuracy as a function of the training epoch computed from 10 trainings.}
	\label{fig:four-graphs}
\end{figure}

To see if HPO improved the model performance, the loss and classification accuracy of the model before and after hypertuning is plotted and compared as a function of the training epoch in figure \ref{fig:four-graphs}.  Comparing these curves shows that the mean validation loss decreased by almost a factor of two (approximately $44\%$) and that the accuracy increased by more than the uncertainty. It is also clear from comparison of figures \ref{fig:loss-before} and \ref{fig:loss-after} as well as of figures \ref{fig:acc-before} and \ref{fig:acc-after} that the training became more stable as a result of hypertuning since the curves exhibit much less volatility after hypertuning, especially in the second half of the training.

\section{Distributed training and hypertuning: synergies across sciences in RAISE}
\label{sec:raise}

\begin{figure}[htb]
	\begin{minipage}{0.5\linewidth}
		\includegraphics[width=1\linewidth]{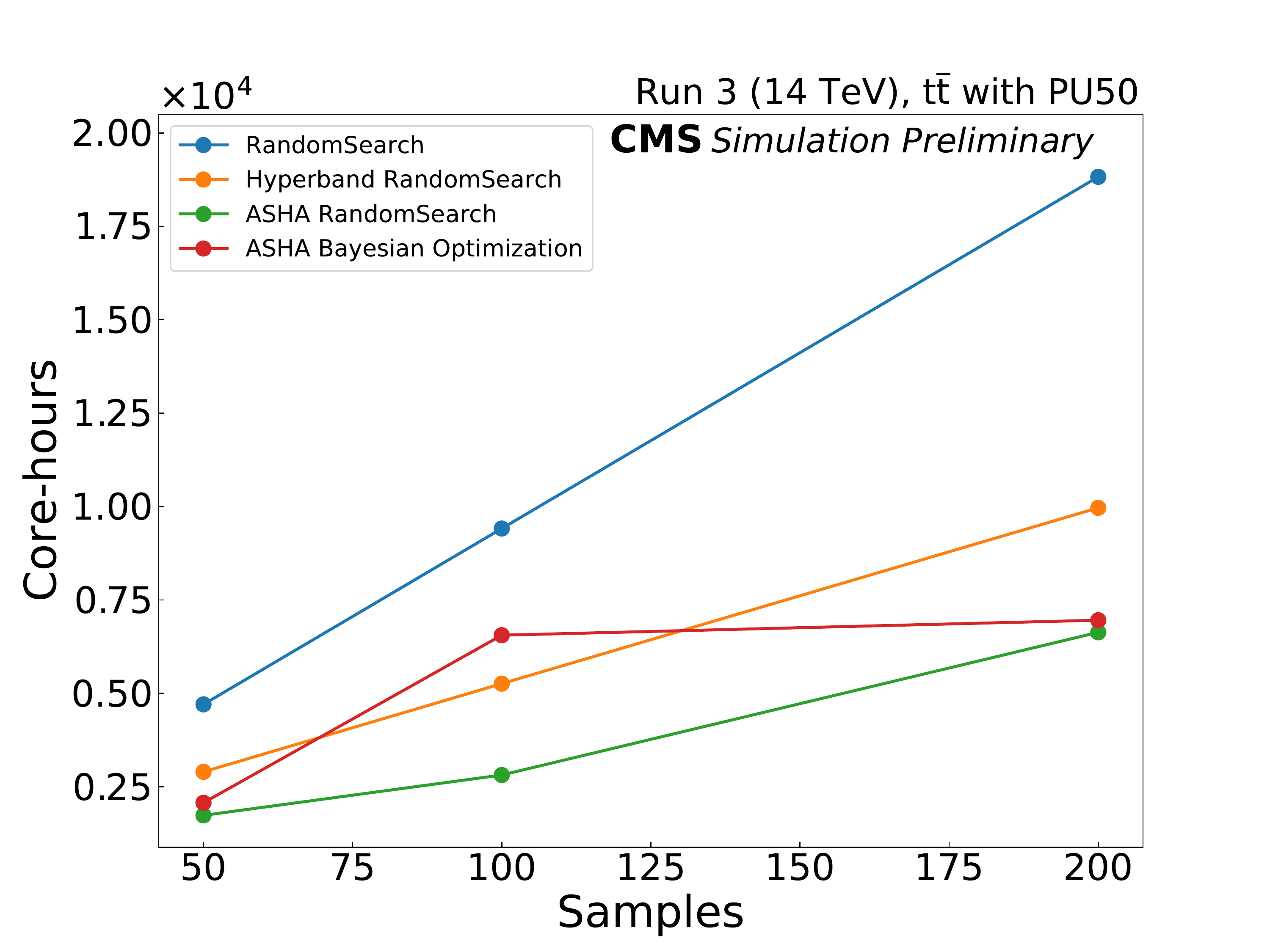}
		\\ \vspace{-7mm}
		\caption{Comparison of hyperparameter search algorithms. Number of core-hours spent versus the number of samples drawn.}
		\label{fig:core-hours-vs-samples}
	\end{minipage}\hspace{1pc}%
	\begin{minipage}{0.5\linewidth}
		\includegraphics[width=1\linewidth]{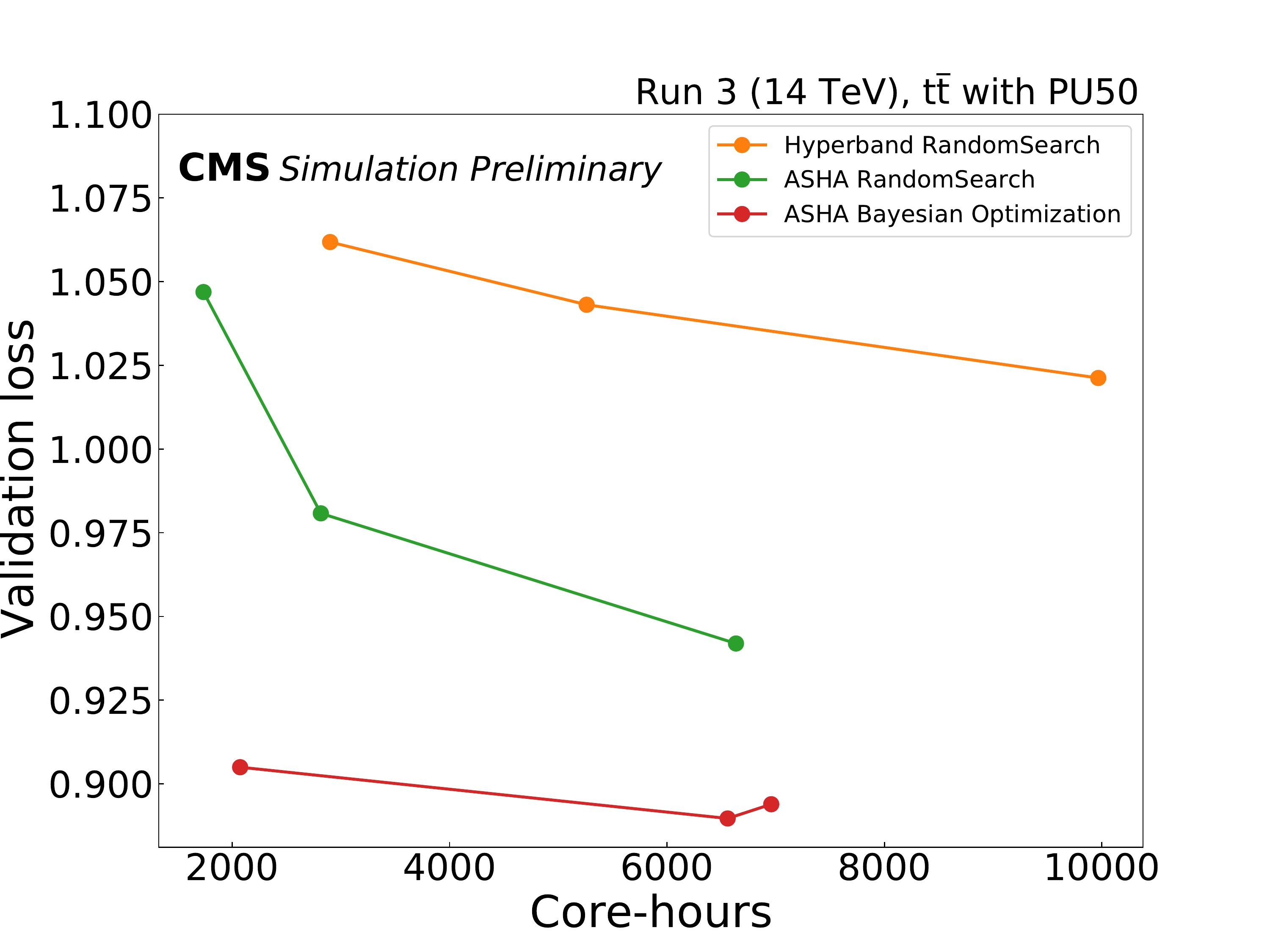}
		\\ \vspace{-7mm}
		\caption{Comparison of hyperparameter search algorithms. Validation loss of the best trial versus number of core-hours spent.}
		\label{fig:val-loss-vs-core-hours}
	\end{minipage} 
\end{figure}

HPO algorithms are model-agnostic in their nature and could be applied in any field of science making use of AI. Hence, the benchmarking of HPO algorithms is of interest for all use cases in CoE RAISE WP4. In light of this, the hypertuning of MLPF was used as an example workflow to benchmark HPO algorithms in Ray Tune by running a variety of them using
four compute nodes. The number of samples drawn were varied and results were analyzed in terms of samples drawn, compute hours spent, best achieved validation loss and best improvement per compute resources spent. The results are presented in figures \ref{fig:core-hours-vs-samples} and \ref{fig:val-loss-vs-core-hours}. Figure \ref{fig:core-hours-vs-samples} shows that both Hyperband and ASHA significantly outperforms random search in terms of core-hours spent per sample. Comparing the runs using ASHA, the combination with Bayesian optimization adds some overhead compared to the combination with random search making the ASHA + random search combination perform best from this perspective. Figure \ref{fig:val-loss-vs-core-hours} gives another point of view, where the validation loss is plotted against the core-hours spent. From this viewpoint, it is clear that ASHA + Bayesian optimization gives the highest improvement per spent core-hour.

\section{Conclusion}
\label{sec:conclusion}

CoE RAISE develops novel, scalable AI methods towards Exascale with use cases from a wide range of sciences and industry. HPO could benefit any data-driven AI-based algorithm and in  the example use case of MLPF,  large-scale distributed hypertuning significantly increased model performance. This would not have been possible without access to HPC resources since it would have taken approximately half a year of continuous hypertuning on a single GPU. Other sciences and use cases in CoE RAISE are also adopting HPC for hypertuning, including the use cases of WP4, within fields such as remote sensing, seismic imaging, defect-free additive manufacturing and sound engineering.

\ack
We thank our colleagues in CoE RAISE, in particular Andreas Lintermann, Morris Riedel, Marcel Aach, Eric Michael Sumner, Eray Inanc, Michael Bresser, Jennifer Lopez Barrilao, Ieva Timrote and Christina Bolanou for helpful discussions and feedback in the course of this work. We also thank our colleagues in the CMS Collaboration, especially Javier Duarte, Farouk Mokhtar, Jieun Yoo, Jean-Roch Vlimant and Maurizio Pierini for their contributions to MLPF.

Eric Wulff was supported by CoE RAISE and Joosep Pata was supported by the Mobilitas Pluss Grant No. MOBTP187 of the Estonian Research Council. The CoE RAISE project has received funding from the European Union’s Horizon 2020 – Research and Innovation Framework Programme H2020-INFRAEDI-2019-1 under grant agreement no. 951733.


\begingroup
\section*{References}
\bibliography{main}
\endgroup

\end{document}